\documentclass{ifacconf}

\usepackage{graphicx}      
\usepackage{natbib}        
\usepackage{enumitem}
\usepackage{amsmath,amssymb,amsfonts}
\usepackage{booktabs}
\usepackage{xcolor} 
\definecolor{red}{rgb}{1,0,0}
\definecolor{blue}{rgb}{0,0,1}
\definecolor{green}{rgb}{0,1,0}
\definecolor{yellow}{rgb}{1,1,0}
\definecolor{orange}{rgb}{1,0.647,0}
\definecolor{gold}{rgb}{1,0.843,0}
\definecolor{purple}{rgb}{0.627,0.125,0.941}
\definecolor{gray}{rgb}{0.745,0.745,0.745}
\definecolor{brown}{rgb}{0.647,0.165,0.165}
\definecolor{navy}{rgb}{0,0,0.502}
\definecolor{pink}{rgb}{1,0.753,0.796}
\definecolor{seagreen}{rgb}{0.18,0.545,0.341}
\definecolor{turquoise}{rgb}{0.251,0.878,0.816}
\definecolor{violet}{rgb}{0.933,0.51,0.933}
\definecolor{darkblue}{rgb}{0,0,0.545}
\definecolor{darkcyan}{rgb}{0,0.545,0.545}
\definecolor{darkgray}{rgb}{0.663,0.663,0.663}
\definecolor{darkgreen}{rgb}{0,0.392,0}
\definecolor{darkmagenta}{rgb}{0.545,0,0.545}
\definecolor{darkorange}{rgb}{1,0.549,0}
\definecolor{darkred}{rgb}{0.545,0,0}
\definecolor{lightblue}{rgb}{0.678,0.847,0.902}
\definecolor{lightcyan}{rgb}{0.878,1,1}
\definecolor{lightgray}{rgb}{0.827,0.827,0.827}
\definecolor{lightgreen}{rgb}{0.565,0.933,0.565}
\definecolor{lightyellow}{rgb}{1,1,0.878}
\definecolor{black}{rgb}{0,0,0}
\definecolor{white}{rgb}{1,1,1}
\definecolor{mblue}{rgb}{0, 0.4470, 0.7410}
\definecolor{mred}{rgb}{0.85, 0.325, 0.098}
\definecolor{morange}{rgb}{0.929, 0.694, 0.125}
\definecolor{mpurple}{rgb}{0.494, 0.184, 0.556}
\definecolor{mgreen}{rgb}{0.466, 0.674, 0.188}
\definecolor{mcyan}{rgb}{0.301, 0.745, 0.933}
\definecolor{mbrown}{rgb}{0.635, 0.078, 0.184}

\usepackage{tikz}
\usetikzlibrary{plotmarks}
\usetikzlibrary{patterns,ipe}
\tikzstyle{ipe stylesheet} = [
  ipe import,
  even odd rule,
  line join=round,
  line cap=butt,
  ipe pen normal/.style={line width=0.4},
  ipe pen heavier/.style={line width=0.8},
  ipe pen fat/.style={line width=1.2},
  ipe pen ultrafat/.style={line width=2},
  ipe pen normal,
  ipe mark normal/.style={ipe mark scale=3},
  ipe mark large/.style={ipe mark scale=5},
  ipe mark small/.style={ipe mark scale=2},
  ipe mark tiny/.style={ipe mark scale=1.1},
  ipe mark normal,
  /pgf/arrow keys/.cd,
  ipe arrow normal/.style={scale=7},
  ipe arrow large/.style={scale=10},
  ipe arrow small/.style={scale=5},
  ipe arrow tiny/.style={scale=3},
  ipe arrow normal,
  /tikz/.cd,
  ipe arrows, 
  <->/.tip = ipe normal,
  ipe dash normal/.style={dash pattern=},
  ipe dash dotted/.style={dash pattern=on 1bp off 3bp},
  ipe dash dashed/.style={dash pattern=on 4bp off 4bp},
  ipe dash dash dotted/.style={dash pattern=on 4bp off 2bp on 1bp off 2bp},
  ipe dash dash dot dotted/.style={dash pattern=on 4bp off 2bp on 1bp off 2bp on 1bp off 2bp},
  ipe dash normal,
  ipe node/.append style={font=\normalsize},
  ipe stretch normal/.style={ipe node stretch=1},
  ipe stretch normal,
  ipe opacity 10/.style={opacity=0.1},
  ipe opacity 30/.style={opacity=0.3},
  ipe opacity 50/.style={opacity=0.5},
  ipe opacity 75/.style={opacity=0.75},
  ipe opacity opaque/.style={opacity=1},
  ipe opacity opaque,
]

\newcommand{\norm}[1]{\left\lVert #1 \right\rVert}
\newcommand{\abs}[1]{\left\lvert #1 \right\rvert}
\newcommand{\T}{T}

\newcommand{\drawlinelegend}[1]{\raisebox{.5ex}{\tikz{\draw[#1, line width=0.4mm] (0,0) -- +(1em, 0);}}}
\newcommand{\drawrectanglelegend}[1]{\raisebox{.0ex}{\tikz[ipe stylesheet]{\filldraw[color=#1!100, fill=#1!15] (0,0) rectangle (2ex, 1ex);}}}

\makeatletter
\g@addto@macro\normalsize{%
}
\makeatother

\usepackage{eso-pic}
\AddToShipoutPictureBG*{%
	\AtPageUpperLeft{%
		\setlength\unitlength{1in}%
		\hspace*{\dimexpr0.5\paperwidth\relax}
		\makebox(0,-2)[c]{\begin{tabular}{c c}
				Learning for Precision Motion of an Interventional X-ray System: \\ Add-on Physics-Guided Neural Network Feedforward Control \\
				To appear in {\em 22$^{nd}$ IFAC World Congress}, Yokohama, Japan, 2023, uploaded to ArXiv \today
		\end{tabular}}
}}
\begin{document}
\begin{frontmatter}

\title{Learning for Precision Motion of an Interventional X-ray System: Add-on Physics-Guided Neural Network Feedforward Control\thanksref{footnoteinfo}}

\thanks[footnoteinfo]{This work is supported by Topconsortia voor Kennis en Innovatie (TKI), and ASML and Philips Engineering Solutions.}

\author[First]{Johan Kon} 
\author[First]{Naomi de Vos}
\author[Second]{Dennis Bruijnen}
\author[Third]{Jeroen van de Wijdeven}
\author[First,Third]{Marcel Heertjes}
\author[First,Fourth]{Tom Oomen} 

\address[First]{Control Systems Technology Group, Departement of Mechanical Engineering, Eindhoven University of Technology, P.O. Box 513, 5600 MB Eindhoven, The Netherlands. e-mail: j.j.kon@tue.nl.}
\address[Second]{Philips Engineering Solutions, Eindhoven, The Netherlands.}
\address[Third]{ASML, Veldhoven, The Netherlands.}
\address[Fourth]{Delft Center for Systems and Control, Delft University of Technology, The Netherlands.}

\begin{abstract}                
Tracking performance of physical-model-based feedforward control for interventional X-ray systems is limited by hard-to-model parasitic nonlinear dynamics, such as cable forces and nonlinear friction. In this paper, these nonlinear dynamics are compensated using a physics-guided neural network (PGNN), consisting of a physical model, embedding prior knowledge of the dynamics, in parallel with a neural network to learn hard-to-model dynamics. To ensure that the neural network learns only unmodelled effects, the neural network output in the subspace spanned by the physical model is regularized via an orthogonal projection-based approach, resulting in complementary physical model and neural network contributions. The PGNN feedforward controller reduces the tracking error of an interventional X-ray system by a factor of 5 compared to an optimally tuned physical model, successfully compensating the unmodeled parasitic dynamics. 
\end{abstract}


\end{frontmatter}
\section{Introduction}
\label{sec:introduction}
Image-guided therapy (IGT) in general, and interventional X-rays specifically, are a key technology in healthcare that directly improve treatment quality by enabling minimally invasive therapies through visualization of patient tissue \citep{Jolesz2014}. The interventional X-ray is one of the main IGT systems and is able to create 3D images of the relevant tissue by combining a sequence of 2D X-ray snapshots \citep{Pelc2014}. It is specifically geared toward use during surgery to generate real-time images of the relevant tissue, enabling small surgical tools as opposed to making large incisions, resulting in faster patient recovery.

Accurate feedforward control \citep{Clayton2009, Butterworth2009} is essential during operation of an interventional X-ray system to guarantee both high imagine quality as well as patient and operator safety. First, accurate feedforward control allows for compensation of the system's dynamics before errors occur, resulting in accurate tracking of the desired setpoint for the imaging sequence, minimizing visual artefacts such as motion blur. Second, mismatch between motor torques predicted by feedforward control and actual applied torques, i.e., the feedback controller contribution, can be used as a basis to distinguish nominal operating conditions from anomalies such as collisions, increasing safety.

Feedforward controllers based on physical models \citep{Devasia2002, Zou2004, 489285} have limited performance due to hard-to-model or unknown dynamics present in an interventional X-ray: dynamics not included in the physical model are not compensated through feedforward control, resulting in reproducible tracking errors. At the same time, these physical models are highly flexible, i.e., result in the same performance for different trajectories \citep{6837472}, and can describe the majority of the dynamics in terms of simple expressions using a few interpretable parameters, such as mass and snap coefficients \citep{Boerlage2003}.

In sharp contrast to a physical-model-based approach, neural network feedforward controllers can compensate all predictable dynamics of any system given a sufficiently rich (recurrent) parametrization \citep{Goodfellow2016, Schafer2006} and have been successfully applied in feedforward control to improve tracking performance \citep{HUNT19921083, Sorensen1999, Otten1997}. However, neural networks are challenging to optimize \citep{80336}, uninterpretable, and lack the ability to extrapolate based on physical prior knowledge \citep{Schoukens2019}. 

Physics-guided neural networks (PGNNs) aim to reconcile the performance of neural networks with the flexibility and interpretability of physical models by explicitly introducing the physics into the structure \citep{7959606} or optimization criterion \citep{2017arXiv171011431K}. These PGNNs can be shown to increase performance over physical-model-based feedforward controllers \citep{Bolderman2021}. Despite their improvement, the contributions of the physical model and neural network are not always well distinguished. Complementarity is obtained by explicitly separating the physical model and neural network contribution through orthogonal projection-based regularization \citep{Kon2022PhysicsGuided}, resulting in physically meaningful model coefficients.


The main contribution of this paper is the development and experimental validation of the PGNN feedforward framework to learn and subsequently compensate the hard-to-model dynamics present in the interventional X-ray. This consists of the following subcontributions.
\begin{enumerate}[label=C\arabic*)]
	\item A PGNN feedforward parametrization for the interventional X-ray system, consisting of a physical model describing its equations of motion and a suitable neural network (Section \ref{sec:physics_guided_neural_network}).
	\item An orthogonal projection-based regularizer to ensure complementarity of the physical model and neural network (Section \ref{sec:orthogonal_projection_based_regularization}).
	\item Experimental validation of the PGNN feedforward controller on an interventional X-ray system, illustrating its superior tracking performance over physical-model-based feedforward control (Section \ref{sec:experimental_validation}).
\end{enumerate}

\section{Control Problem}
\label{sec:problem_formulation}
\subsection{Interventional X-ray System}
\begin{figure}[t]
	\centering
	\includegraphics[width=\linewidth]{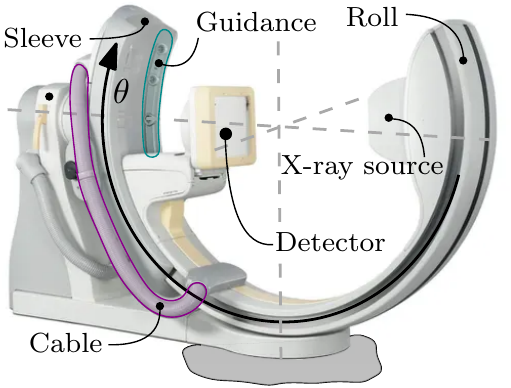}
	\caption{Interventional X-ray system with roll axis with orientation $\theta$ positioning the X-ray source and detector. The configuration-dependent cable forces (\protect \drawrectanglelegend{darkmagenta}) and friction characteristics of the guidance (\protect \drawrectanglelegend{darkcyan}) limit the effectiveness of physical-model-based feedforward control.}
	\label{fig:CLEA}
\end{figure}

The considered interventional X-ray system is depicted in Fig. \ref{fig:CLEA}. While it has 3 degrees of freedom, this paper considers only the roll axis. The roll axis body rotates in a roller-based guidance attached to the sleeve, thus positioning the X-ray source and detector in this dimension. It is driven by a permanent magnet DC motor and an amplifier with a maximum input of 5 [V] through a transmission consisting of gears and belts. The rotation is measured using an incremental encoder with an effective resolution of $0.0119$ [deg]. Encoder measurements and power for the X-ray source are supplied through a cable at the side of the setup, which is acting as a dynamic link. The software runs on a Speedgoat system with $T_s= \frac{1}{500}$ [s].

This mechanical design, motivated and constrained by the use around medical personnel, introduces the following hard-to-model nonlinear dynamics.
\begin{enumerate}
	\item The mass distribution is unbalanced, resulting in configuration-dependent gravitational forces.
	\item The cable acts as a configuration-dependent inertia, and is tensioned for large negative $\theta$, acting as a one-sided spring.
	\item The friction characteristics in the guidance depend on the normal forces acting on the contact surface of the rollers, and are thus configuration-dependent. 
\end{enumerate} 
The configuration-dependent gravitational forces can be quantitatively captured by a physical model. However, the exact way these parasitic cable and friction forces depend on the configuration $\theta$ is exceptionally hard to model from a physics-based perspective, limiting the tracking performance of physical-model-based feedforward control.

\subsection{Control Approach}
\begin{figure}[t]
	\centering
	\includegraphics[width=\linewidth]{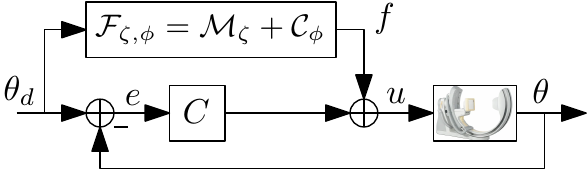}
	\caption{Two-degree-of-freedom control configuration for roll axis of an interventional X-ray system with feedback controller $C$ and feedforward controller $\mathcal{F}_{\zeta,\phi}$ consisting of physical model $\mathcal{M}_\zeta$ and neural network $\mathcal{C}_\phi$. }
	\label{fig:control_setup}
\end{figure}

The main dynamics of the interventional X-ray system, i.e., the configuration-dependent gravitational forces, can be described by physics, but the cable and friction forces are only qualitatively understood and hard to model. Therefore, the physical model $\mathcal{M}_\zeta$ with parameters $\zeta$ is complemented by a neural network $\mathcal{C}_\phi$ to learn these hard-to-model dynamics, resulting in a parallel PGNN feedforward controller $\mathcal{F}_{\zeta,\phi}$, see Fig. \ref{fig:control_setup}. Additionally, a simple PD feedback controller is employed to compensate both unknown external disturbances as well as dynamics uncompensated by the PGNN feedforward controller.

The aim of this paper is to learn the parameters $\zeta,\phi$ of the PGNN feedforward controller (defined in the next section) that compensate the hard-to-model cable and friction force, thereby increasing tracking performance, in such a way that $\mathcal{M}_\zeta$ stays interpretable and $\mathcal{C}_\phi$ increases performance by learning only unmodeled dynamics, i.e., while ensuring complementarity. In this paper, an inverse system identification approach is taken, in which $\zeta,\phi$ are learned based on a dataset $\mathcal{D}$ describing the system's dynamics consisting of inputs $\hat{u}(k)$ corresponding to outputs $\theta(k)$ with $k=1,\ldots,N$ that has been obtained in advance, e.g., feedback or iterative learning control data. 

\section{Physics-Guided Neural Network}
\label{sec:physics_guided_neural_network}%
\subsection{Parallel Feedforward Parametrization}
\begin{figure}[t]
	\centering
	\includegraphics[width=\linewidth]{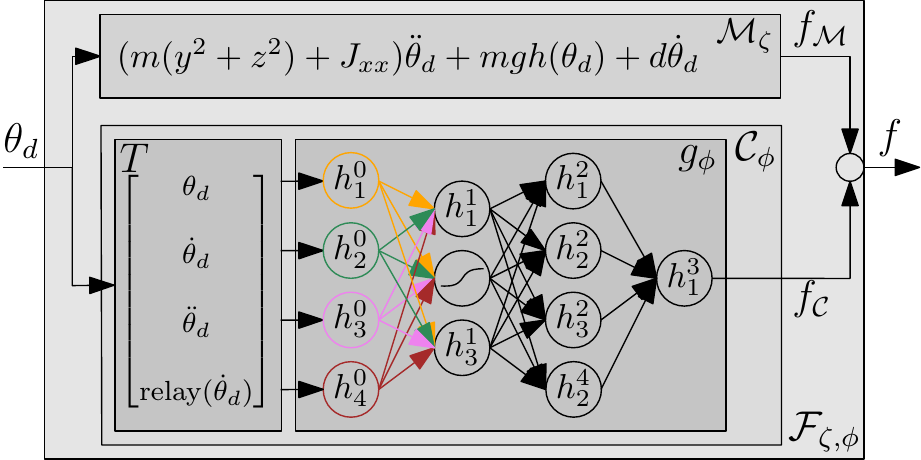}
	\caption{PGNN feedforward controller consisting of physical-model-based feedforward controller $\mathcal{M}_\zeta$ with parameters $\zeta$ and neural network feedforward controller $\mathcal{C}_\phi$. $\mathcal{C}_\phi$ consists of a feedforward neural network $g_\phi$ and physics-guided input $T(\theta_d(k))$, here with $L=3$ hidden layers of size $4, 3, 4$, and a linear output layer.  }
	\label{fig:PGNN_feedforward}
\end{figure}

To compensate hard-to-model dynamics, the feedforward controller $\mathcal{F}_{\zeta,\phi}$ is parametrized as the parallel combination of a physical model $\mathcal{M}_\zeta$ and complementary neural network $\mathcal{C}_\phi$ according to
\begin{equation}
	f(k) = \mathcal{F}_{\zeta,\phi}(\theta_d(k)) = \mathcal{M}_\zeta(\theta_d(k)) + \mathcal{C}_\phi(\theta_d(k)),
	\label{eq:parallel_feedforward}
\end{equation}
and is visualized in Fig. \ref{fig:PGNN_feedforward}.
The physical model $\mathcal{M}_\zeta$ with parameters $\zeta$ represents prior knowledge of the physical process through encapsulating quantitatively known dynamics. The physical model is complemented by a universal function approximator $\mathcal{C}_\phi$ with parameters $\phi$, here chosen as a neural network, such that it can learn dynamics not included in the physical model.

\subsection{Physical Model}
The physical model $\mathcal{M}_\zeta$ constitutes an equation of motion describing the relation between the generalized coordinate $\theta$ and the input $u$ derived from first-principles modeling \citep{deVos2022}. More specifically, assuming that inputs are confined within actuator limits and ignoring drivetrain flexibilities, the equation of motion for $\theta$ is given by
\begin{equation}
	u = M \ddot{\theta} + H(\theta,\dot{\theta}) + d\dot{\theta}.
	\label{eq:physical_model}
\end{equation}
Here, $d \in \mathbb{R}_{\geq 0}$ is the viscous damping coefficient,
\begin{equation}
	M = m (y^2 + z^2) + J_{xx} \in \mathbb{R}_{\geq 0},
\end{equation}
the inertia of the roll axis, and
\begin{equation}
	H(\theta,\dot{\theta}) = m g (y \cos(\theta) - z \sin(\theta))\cos(\phi) \in \mathbb{R},
\end{equation}
the gravity contribution. Coordinates $y,z\in\mathbb{R}$ represent the offset of the center of mass with respect to the point of rotation, and $\phi$ the known orientation of the roll axis out of the vertical plane.

The required feedforward signal can be obtained by evaluating the right-hand-side for a given reference $\theta_d$ and discrete-time derivatives $\dot{\theta}_d$. Thus, the physical-model-based feedforward controller $\mathcal{M}_\zeta: \theta_d \rightarrow f_\mathcal{M}$ is given by
\begin{equation}
	f_\mathcal{M}(k) = M \ddot{\theta}_d(k) + H(\theta_d(k),\dot{\theta}_d(k)) + d\dot{\theta}_d(k),
	\label{eq:f_M}
\end{equation}
with physical parameters 
\begin{equation}
	\zeta = \begin{bmatrix} m & J_{xx}& d & y & z	\end{bmatrix}^\T \in \mathbb{R}^{N_\zeta}.
\end{equation}

\subsection{Neural Network}

The neural network $\mathcal{C}_\phi$ consists of a feedforward neural network (FNN) acting on a physics-guided input vector. By the universal approximation theorem \citep{Goodfellow2016}, $\mathcal{C}_\phi$ can represent any continuous function on a bounded interval up to arbitrary precision, such that it can learn hard-to-model dynamics, such as cable forces, from data.

More specifically, the neural network feedforward controller $\mathcal{C}_\phi: \theta_d \rightarrow f_\mathcal{C} $ is defined as
\begin{equation}
	f_\mathcal{C}(k) = g_\phi(T(\theta_d(k))).
	\label{eq:f_C}
\end{equation}
The FNN $g_\phi: T(\theta_d(k)) = x \rightarrow f_\mathcal{C}(k)$ is given by a fully connected multilayer perceptron (MLP) of $L$ hidden layers and a linear output layer without bias, i.e.,
\begin{align}
	h^l(x) &= x	 & \textrm{if}\ \ & l = 0 \nonumber \\
	h^l(x) &= \sigma \left( W^{l-1} h^{l-1}(x) + b^l \right) & \textrm{if}\ \  & l = {1, \ldots, L} \nonumber \\
	g_{\phi}(x) &= W^l h^{l}(x) & \textrm{if}\ \ & l = L, \label{eq:FNN}
\end{align}
in which $W^l,b^l$ are appropriately sized weight matrices and bias vectors defining affine mappings, and $\sigma$ is an element-wise activation function acting on this affine mapping, such as a sigmoid, hyperbolic tangent ($\tanh$) or rectified linear unit (ReLU). The full MLP $g_\phi$ then is the repeated application of affine and nonlinear transformations. 

The physics-guided input transformation $T$ encodes prior qualitative knowledge on the hard-to-model dynamics and is defined as
\begin{equation}
	T(\theta_d(k)) = \begin{bmatrix} \theta_d(k) & \dot{\theta}_d(k) & \ddot{\theta}_d(k) & \mathrm{relay}(\dot{\theta}_d(k))	\end{bmatrix}^\T,
\end{equation}
with 
\begin{equation}
\mathrm{relay}(x(k)) = \begin{cases}
1 & \textrm{if } x(k) > 0 \\
-1 & \textrm{if } x(k) < 0 \\
\mathrm{relay}(x(k-1)) & \textrm{if } x(k) = 0,
\end{cases}
\label{eq:relay}
\end{equation}
and $\mathrm{relay}(x(0)) = 0$.
It transforms $\theta_d$ such that the input to $g_\phi$ contains all relevant physical quantities to make predictions about the required force.
For example, $T(\theta_d(k))$  contains the reference velocity $\dot{\theta}_d$ to learn complex friction characteristics, and a relay of the velocity to approximate hysteresis characteristics without requiring a recurrent network architecture, thereby greatly simplifying optimization.

\section{Separating Physical Model and Neural Network Contributions}
\label{sec:orthogonal_projection_based_regularization}
The neural network $\mathcal{C}_\phi$ is also able to learn and compensate dynamics included in the physical model $\mathcal{M}_\zeta$ and thus there exists multiple parameters $\zeta,\phi$ that produce the same input-output behaviour of $\mathcal{F}_{\zeta,\phi}$, i.e., the PGNN parametrization \eqref{eq:parallel_feedforward} is unidentifiable. Consequently, the physical model and neural network need not be complementary. By regularizing the neural network contribution $f_\mathcal{C}$ in the output space of the physical model, a specific solution on this manifold is selected, namely the one in which modeled dynamics are compensated by the physical model component and not by the neural network.

\subsection{Least-Squares And Consequences of Unidentifiability}
To illustrate unidentifiability, the least-squares cost function is decomposed in two orthogonal subspaces, one spanned by the physical model (locally, given some fixed parameter) and one as its orthogonal complement, from which it is apparent that the neural network can have a contribution in the subspace spanned by the physical model, resulting in unidentifiability.

Consider the least-squares criterion $J_{LS}$ penalizing residuals between the output of $\mathcal{F}_{\zeta,\phi}(\theta(k))$ and the required input $\hat{u}(k)$ for this $\theta(k)$, i.e., 
\begin{equation}
J_{LS} = \sum_{k=1}^N \norm{\hat{u}(k) - (f_\mathcal{M}(k) + f_\mathcal{C}(k))}_2^2 \in \mathbb{R}_{\geq 0}.
\label{eq:LS_costs}
\end{equation}
To decompose this cost function into orthogonal subspaces, the physical model parameters are split into two subsets $\zeta_l$, $\zeta_n$, such that, given $\zeta_n$, $f_\mathcal{M}$ is linear in $\zeta_l$, i.e.,
\begin{equation}
	\begin{aligned}
		\zeta_l &= \begin{bmatrix} m & J_{xx} & d \end{bmatrix}^\T \in \mathbb{R}^{N_{\zeta_l}} & & & \zeta_n &= \begin{bmatrix} y & z	\end{bmatrix} \in \mathbb{R}^{N_{\zeta_n}}.
	\end{aligned}
\end{equation}
Then, the physical model response can be written as
\begin{align}
f_\mathcal{M}(k) &= \begin{bmatrix} x_{1,\zeta_n}(\theta_d(k)) & x_2(\theta_d(k)) & x_3(\theta_d(k)) \end{bmatrix} \begin{bmatrix} m & J_{xx} & d \end{bmatrix}^\T \nonumber \\
&= X_{\zeta_n}(\theta_d(k)) \zeta_l
\end{align}
with ($\zeta_n$-dependent) basis functions
\begin{equation}
	\begin{gathered}
		x_{1,\zeta_n}(\theta) = (y^2 + z^2) \ddot{\theta} + g(y\cos(\theta) - z \sin(\theta)) \cos(\phi) \\
		\begin{aligned}
		x_2(\theta) &= \ddot{\theta} & & & x_3(\theta) &= \dot{\theta}.
		\end{aligned}
	\end{gathered}
\end{equation}
Now, represent finite-time signal $\hat{u}(k)$, $k=1,\ldots,N$ as a vector, i.e., 
\begin{equation}
	\underline{\hat{u}} = \begin{bmatrix} \hat{u}(1) & \hat{u}(2) & \hdots & \hat{u}(N) \end{bmatrix}^\T \in \mathbb{R}^N,
\end{equation}
and similarly for $\underline{\theta}_d$, $\underline{f}_\mathcal{M}$, $\underline{f}_\mathcal{C}$. Then, $\underline{f}_\mathcal{M}$ can be written as 
\begin{equation}
	\underline{f}_\mathcal{M} = X_{\zeta_n}(\underline{\theta}_d) \zeta_l,
	\label{eq:physical_model_pseudo_LIP}
\end{equation}
with finite-time basis function matrix
\begin{equation}
	X_{\zeta_n}(\underline{\theta}_d) = \begin{bmatrix} X_{\zeta_n}^\T (\theta_d(1)) & \hdots	& X_{\zeta_n}^\T (\theta_d(N)) \end{bmatrix}^\T \in \mathbb{R}^{N \times N_{\zeta_l}}.
\end{equation}
Based on \eqref{eq:physical_model_pseudo_LIP}, the output space of the physical model for any $\zeta_l$, given $\zeta_n$, is formed by the image of the basis function matrix $X_{\zeta_n}$ evaluated for the data $\underline{\theta}_d$, and can be represented through, e.g., a singular value decomposition (SVD). More specifically, consider the SVD of $X_{\zeta_n}$ as
\begin{equation}
	X_{\zeta_n}(\underline{\theta}_d) = \begin{bmatrix} U_{1, \zeta_n} & U_{2, \zeta_n} \end{bmatrix} \begin{bmatrix} \Sigma_{\zeta_n} & 0 \\ 0 & 0\end{bmatrix} \begin{bmatrix} V_{1, \zeta_n}^T \\ V_{2, \zeta_n}^\T	\end{bmatrix},
\end{equation}
with $\Sigma_{\zeta_n} \in \mathbb{R}^{r \times r}$, $r = \textrm{rank}(X_{\zeta_n})$, and $U_{1, \zeta_n} \in \mathbb{R}^{N \times r}$, $U_{2, \zeta_n} \in \mathbb{R}^{N \times N-r}$
such that $\begin{bmatrix} U_{1, \zeta_n} & U_{2, \zeta_n} \end{bmatrix}$ is an orthonormal matrix, i.e., 
\begin{equation}
	\begin{aligned}
		U_{1, \zeta_n}^\T U_{1, \zeta_n} &= I_{r} & U_{2, \zeta_n}^\T U_{1, \zeta_n} &= 0 & U_{2, \zeta_n}^\T U_{2, \zeta_n} = I_{N-r}.
	\end{aligned}
\end{equation}
Then, $U_{1, \zeta_n}$ forms a basis for the image of $X_{\zeta_n}$, i.e., the output $\underline{f}_\mathcal{M}$ of the physical model for any $\zeta_l$, given $\zeta_n$, lies in the subspace spanned by $U_{1, \zeta_n}$.

Consider again $J_{LS}$ in \eqref{eq:LS_costs}, which with above finite-time notation can also be written as
\begin{equation}
J_{LS} = \norm{\underline{\hat{u}} - X_{\zeta_n}(\underline{\theta}_d) \zeta_l - g_\phi(T(\underline{\theta}_d))}_2^2.
\label{eq:J_LS_lifted}
\end{equation} 
Given basis $U_{1, \zeta_n}$ of the physical model output space, \eqref{eq:J_LS_lifted} is decomposed into $U_{1, \zeta_n}$ and complement $U_{2, \zeta_n}$ as \citep{Kon2022PhysicsGuided}
\begin{equation}
	J_{LS} = \norm{
	\begin{bmatrix}
	U_{1, \zeta_n}^\T \underline{\hat{u}} \\
	U_{2, \zeta_n}^\T \underline{\hat{u}}
	\end{bmatrix}
	-
	\begin{bmatrix}
	\Sigma_{\zeta_n} V_{1, \zeta_n}^\T \zeta_l + U_{1, \zeta_n}^\T g_\phi(T(\underline{\theta}_d)) \\
	U_{2, \zeta_n}^\T g_\phi(T(\underline{\theta}_d))
	\end{bmatrix}
	}_2^2.
	\label{eq:J_LS_decomposed}
\end{equation}
in which it is used that i) multiplying by $U_{1, \zeta_n}$ does not change the norm, ii) $X_{\zeta_n}(\underline{\theta}_d) = U_{1, \zeta_n} \Sigma_{\zeta_n} V_{1, \zeta_n}^\T$, iii) $U_{2, \zeta_n}^\T U_{1, \zeta_n}^\T=0$, and iv) $U_{1, \zeta_n} U_{1, \zeta_n}^\T + U_{2, \zeta_n}^\T U_{2, \zeta_n} = I$. In the decomposed cost function \eqref{eq:J_LS_decomposed}, $U_{1, \zeta_n}^\T g_\phi(T(\underline{\theta}_d))$ represents the neural network contribution in the subspace spanned by the physical model, such that dynamics can be freely interchanged as long as $g_\phi(T(\underline{\theta}_d))$ can have a contribution in $U_{1, \zeta_n}^\T$, revealing the unidentifiability. 

\subsection{Ensuring Complementarity through Regularization}
To ensure that the physical model learns all dynamics that fit in \eqref{eq:f_M}, the neural network contribution \eqref{eq:f_C} in the subspace of the physical model $U_{1, \zeta_n}$ is penalized through orthogonal projection-based regularization, ensuring complementarity between the physical model and neural network.

More specifically, the orthogonal projection-based regularization (OP-regularization) is defined as
\begin{equation}
R(\phi) = \norm{U_{1, \zeta_n^0}^\T g_\phi(T(\underline{\theta}_d))}_2^2 \in \mathbb{R}_{\geq 0},
\label{eq:OP_regularization}
\end{equation}
where $\zeta_n^0$ is an initial estimate of $\zeta_n$, obtained here from the best physical model fit. This regularization penalizes neural network contributions $g_\phi(T(\underline{\theta}_d))$ in the output space of the physical model $U_{1, \zeta_n^0}$, such that the neural network focuses on learning unmodeled dynamics in $U_{2, \zeta_n^0}$. As such, regularization \eqref{eq:OP_regularization} can be seen as targeted $L_2$-regularization, where only the weight directions generating a contribution in $U_{1, \zeta_n^0}$ are shrunk.

The feedforward parametrization \eqref{eq:parallel_feedforward} is then optimized according to
\begin{equation}
	\min_{\zeta, \phi} J_{OP} = \min_{\zeta, \phi} J_{LS} + \lambda R(\phi),
	\label{eq:J_OP}
\end{equation}
with user-defined regularization parameter $\lambda \in \mathbb{R}_{\geq 0}$ creating a spectrum between the least-squares case for $\lambda = 0$ and full orthogonality for $\lambda \rightarrow \infty$. Since $g_\phi$ only has finitely many degrees of freedom, in practice there exists a trade-off between not having a contribution in $U_{1, \zeta_n^0}$ and capturing unmodeled dynamics in $U_{2, \zeta_n^0}$, and $\lambda = 10^{-1}$ works well for most problems.

Although $\zeta_n$ is also updated during optimization of \eqref{eq:J_OP}, thus changing $U_{1, \zeta_n}$, recursively updating $U_{1, \zeta_n}$ in \eqref{eq:OP_regularization} to more accurately match the true physical model output space does not yield different results from a fixed initialization in practice: the best physical model fit provides a good enough approximation to ensure complementarity.

\section{Performance Improvement on an Interventional X-ray System}
\label{sec:experimental_validation}
The OP-regularized PGNN is validated on the interventional X-ray setup described in Section \ref{sec:problem_formulation}. It is shown that the PGNN improves the tracking performance by a factor of 5 (Section \ref{subsec:performance_increase}), and that the physical model and neural network contributions are complementary due to OP-regularization (Section \ref{subsec:complementarity}).

\subsection{Hyperparameters and Optimization Details}
An input-output dataset consisting of input $u$ and corresponding plant output $\theta$ for nominal operating conditions of the interventional X-ray is generated using feedback control, and split into 80$\%$ train, 10$\%$ test and 10$\%$ validation. Based on a hyperparameter study, the neural network has $L=2$ hidden layers with 30 neurons each and $\tanh$ activation functions. The PGNN is trained with OP-regularization using ADAM \citep{Kingma2015} with minibatching and early stopping based on the validation set with a patience of 5 minibatches. The physical model parameters are initialized corresponding to the best approximation using only the physical model. The trained feedforward controllers are evaluated based on the root-mean-square norm $\textrm{RMS}(s) = \textrm{sqrt} (N^{-1}\sum_{k=1}^N s(k)^2)$, mean absolute norm $\textrm{MA}(s) = N^{-1} \sum_{k=1}^N \abs{s(k)}$ and max absolute norm $\norm{s}_\infty = \max_{k=1}^N \abs{s(k)}$.

\subsection{Performance Improvement over Physical-Model-Based Feedforward Control}
\label{subsec:performance_increase}
\begin{figure}[b]
	\centering
	\includegraphics[]{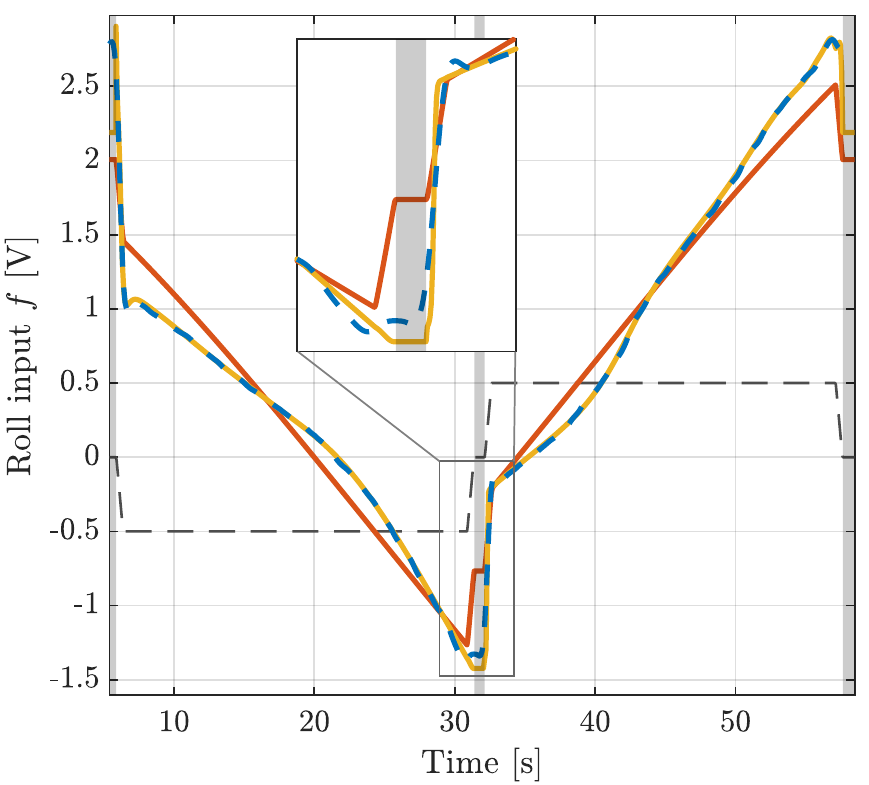}
	\caption{The PGNN feedforward controller generates input (\protect \drawlinelegend{morange}) that matches the required input $\hat{u}$ (\protect \drawlinelegend{dashed, mblue}), learning the friction and cable forces that change as a function of the configuration (scaled velocity \protect \drawlinelegend{dashed,black}). In contrast, the physical-model-based feedforward controller (\protect \drawlinelegend{mred}) does not include these hard-to-model phenomena, instead generating an input with a similar average slope, resulting in predictable residuals.}
	\label{fig:f_dataset_roll}
\end{figure}
The PGNN feedforward parametrization \eqref{eq:parallel_feedforward} significantly improves the performance over a purely physical-model-based approach, both in terms of matching the required input $\hat{u}$ from a dataset, as well as decreasing tracking errors during realtime evaluation, as detailed next.

\subsubsection{Performance on Dataset}
Fig. \ref{fig:f_dataset_roll} shows the generated input $f$ for the PGNN and a purely physical-model-based approach, and compares it to the required input $\hat{u}$. The increased match between $f$ and $\hat{u}$ obtained by learning the configuration-dependent friction and cable characteristics is substantiated by Table \ref{table:f_residual_norm}, illustrating that the PGNN has a of factor 10 smaller input residuals. 

\subsubsection{Realtime Performance}
\begin{figure}[b]
	\centering
	\includegraphics[]{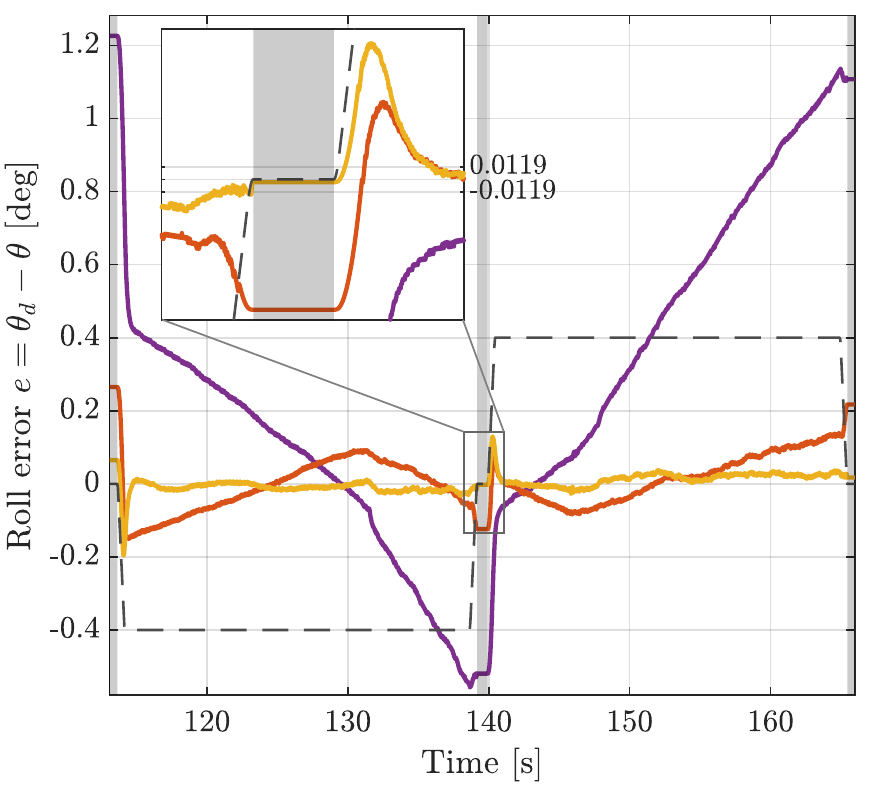}
\caption{The PGNN feedforward controller (\protect \drawlinelegend{morange}) compensates almost all dynamics, resulting in a tracking error of a few encoder counts (indicated in the inset). In contrast, the physical-model-based feedforward controller (\protect \drawlinelegend{mred}) improves upon the feedback only case (\protect \drawlinelegend{mpurple}), but still contains predictable errors from uncompensated dynamics. All approaches suffer from transient errors after stationary periods (\protect \drawrectanglelegend{darkgray}) with zero velocity reference (\protect \drawlinelegend{dashed,black}), potentially caused by stick-slip behaviour.}	
	\label{fig:e_realtime_roll}
\end{figure}
Fig. \ref{fig:e_realtime_roll} compares the tracking errors of the PGNN, purely physical-model-based, and no feedforward controller for a trajectory different from the training data, but with similar maximum velocity and acceleration. The reduction in tracking error due to the inclusion of a neural network is quantified by Table \ref{table:e_realtime_norm_roll} which summarizes the performance norms of the tracking error. The PGNN feedforward controller improves the tracking error by a factor of 5 compared to a physical-model-based approach in terms of MA and RMS norm. The Inf norm is unaltered, as both approaches suffer from transients after stationary periods that cause these errors.

\begin{table}[h]
\centering
\caption{Error norms [deg] for roll axis.}
\label{table:e_realtime_norm_roll}
\begin{tabular}{cccc}
\toprule
 & MA($e$) & RMS($e$) & $\norm{e}_\infty$ \\
\midrule
Feedback & 0.402 & 0.522 & 1.401 \\
Physical model & 0.095 & 0.117 & \textbf{0.279} \\
PGNN & \textbf{0.020} & \textbf{0.029} & \textbf{0.269} \\
\bottomrule
\end{tabular}
\end{table}

\subsection{Complementarity of Neural Network}
\label{subsec:complementarity}
\begin{figure}[t]
	\centering
	\includegraphics[]{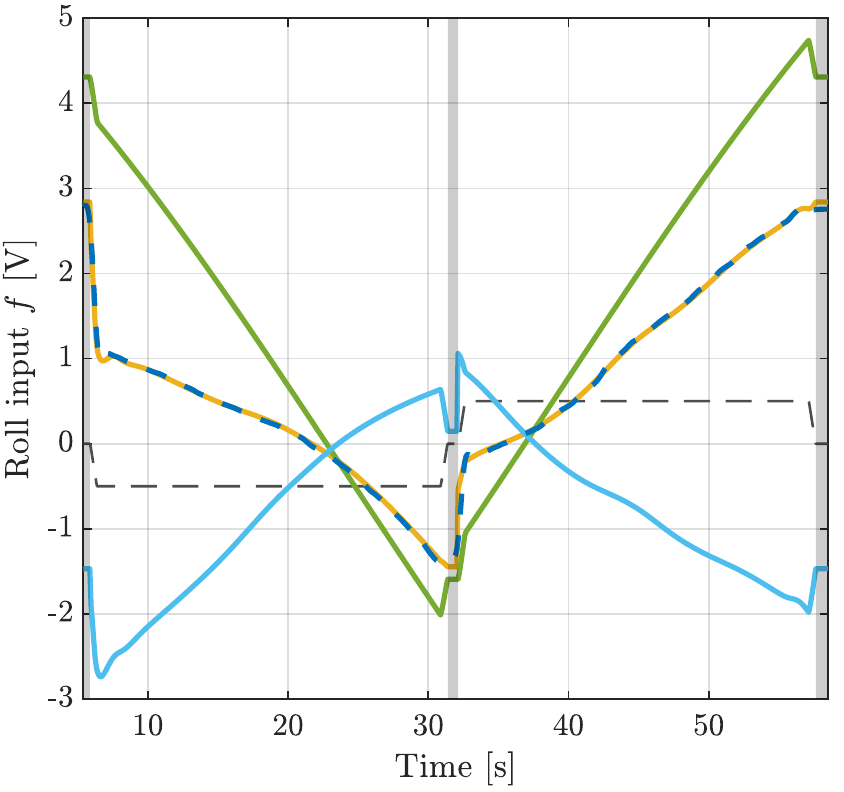}
		\includegraphics[]{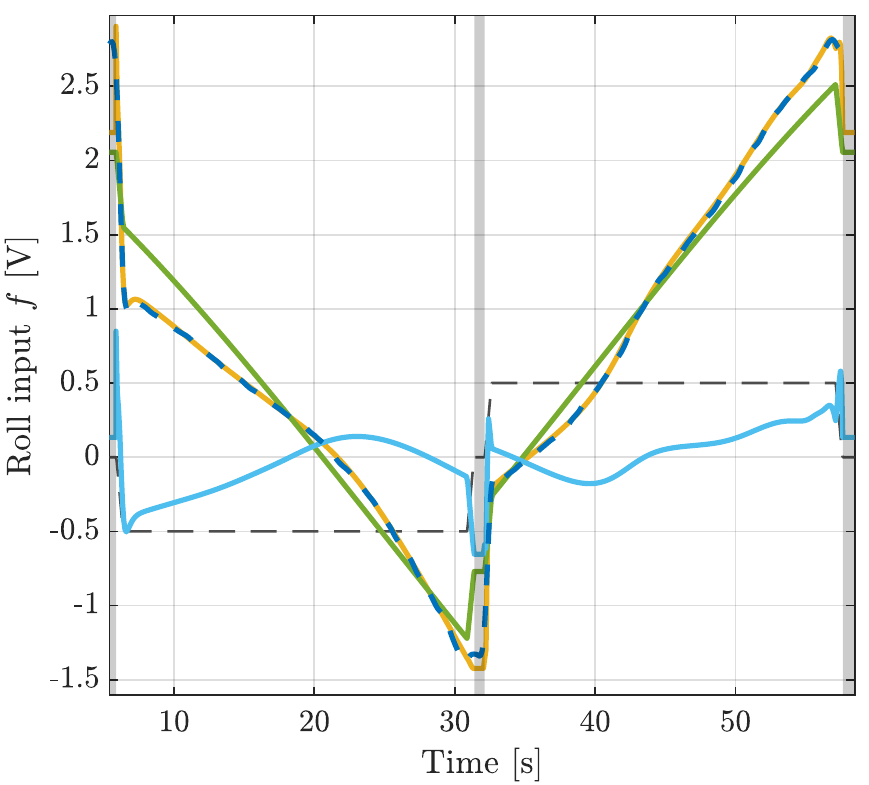}
	\caption{The PGNN feedforward parametrization $\mathcal{F}_{\zeta,\phi}$ is unidentifiable, i.e., there exists multiple parameter sets $\zeta,\phi$ resulting in different physical model (\protect \drawlinelegend{mgreen}) and neural network (\protect \drawlinelegend{mcyan}) contributions that together (\protect \drawlinelegend{morange}) generate the required input $\hat{u}$ (\protect \drawlinelegend{mblue, dashed}). Consequently, training with a least-square criterion (upper) can result in non-complementary contributions and an uninterpretable physical model. In contrast, training with orthogonal projection-based regularization (lower) results in complementary contributions, such that the physical model component remains interpretable and can be used as a baseline. (\protect \drawlinelegend{dashed,black}) represents the scaled velocity reference.}
	\label{fig:f_LS_hybrid_components}
\end{figure}

In addition to improving performance, the orthogonal projection-based regularization \eqref{eq:OP_regularization} also ensures complementarity between the physical model and neural network in PGNN parametrization \eqref{eq:parallel_feedforward}. Fig. \ref{fig:f_LS_hybrid_components} shows the physical model and neural network component of the PGNN for the same interval as Fig. \ref{fig:f_dataset_roll}, for both a PGNN trained with $J_{LS}$ in \eqref{eq:LS_costs} (default PGNN), and one trained with OP-regularized criterion $J_{OP}$ in \eqref{eq:J_OP} (PGNN-OP), from which it follows that a least-squares criterion indeed can result in non-complementary contributions, resulting in an uninterpretable physical model that cannot be used as a baseline.

Table \ref{table:f_residual_norm} shows that both the default PGNN and PGNN-OP can generate the required input $\hat{u}$ up to the same accuracy, but the PGNN-OP requires significantly less neural network contribution due to the regularization enforcing complementarity.

\begin{table}[h]
\centering
\caption{Residual $\varepsilon = \hat{u} - f$ norms [V].}
\label{table:f_residual_norm}
\begin{tabular}{ccccc}
\toprule
 & MA($\varepsilon$) & RMS($\varepsilon$) & $\norm{\varepsilon}_\infty$ & RMS($f_{\mathcal{C}}$) \\
\midrule
Physical model & 0.262 & 0.321 & 1.131 & 0.000 \\
PGNN-OP & 0.016 & 0.038 & 0.842 & 0.325 \\
Default PGNN & 0.020 & 0.046 & 0.711 & 1.410 \\
\bottomrule
\end{tabular}
\end{table}

\section{Conclusion}
\label{sec:conclusion}
Physics-guided neural networks are used to learn and subsequently compensate hard-to-model dynamics such as cable forces and configuration-dependent friction characteristics in an interventional X-ray setup. In this PGNN, the nonlinear equation of motion of one of the X-ray's axis is complemented by a feedforward neural network. The orthogonal projection-based regularizer for this physical model is derived and used during simultaneous optimization to successfully ensure complementarity of the two components. Application of the PGNN results in a factor 5 improved tracking error while maintaining an interpretable physical model that can be used as a baseline. 


\bibliography{../../Literature/Library.bib} 
                                                 
\appendix

\end{document}